%% file: main.tex
\documentclass[preprint,5p,times,number,sort&compress]{elsarticle}
\usepackage{amssymb}
\usepackage{amsmath}
\usepackage{siunitx}
\usepackage{lineno}

\usepackage{float} 
\journal{Matter}

\graphicspath{{figures/}}

\usepackage{xspace}
\NewDocumentCommand{\aimdl}{}{AIMD\nobreakdash-L\xspace}
\NewDocumentCommand{\xray}{}{X\nobreakdash-ray\xspace}

\begin{document}

\begin{frontmatter}


%%            country={USA}}

\title{AIMD-L: An automated laboratory for high-throughput characterization  \\ of structural materials for extreme environments} %% Article title

\author[mse,me,hemi]{Todd C.~Hufnagel}
\author[hemi]{Pranav Addepalli}
\author[lcsr,hemi]{Anuruddha Bhattacharjee}
\author[mse,hemi]{Rohit Berlia}
\author[me,mse,hemi]{Jaafar~El-Awady}
\author[hemi]{David~Elbert}
\author[case,hemi]{Lori~Graham-Brady}
\author[me,lcsr,hemi]{Axel~Krieger}
\author[hemi]{Harichandana Neralla}
\author[hemi]{T. Joseph Nkansah-Mahaney}
\author[me,hemi]{Mostafa M. Omar}
\author[mse,hemi]{Hyun Sang Park}
\author[me,mse,hemi]{K.T.~Ramesh}
\author[hemi]{Matthew Shaeffer}
\author[hemi]{Eric~Walker}
\author[me,hemi]{Piyush Wanchoo}
\author[mse,me,hemi]{Timothy~P.~Weihs}

\affiliation[mse]{organization={Department of Materials Science and Engineering, Johns Hopkins University},
             addressline={},
             city={Baltimore},
             state={Maryland},
             postcode={21218},
              country={USA}}
\affiliation[me]{organization={Department of Mechanical Engineering, Johns Hopkins University},
             addressline={},
             city={Baltimore},
             postcode={21218},
             state={Maryland},
             country={USA}}
\affiliation[case]{organization={Department of Civil and Systems Engineering, Johns Hopkins University},[3]
             addressline={},
             city={Baltimore},
             state={Maryland},
             postcode={21218},
             country={USA}}

\affiliation[lcsr]{organization={Laboratory for Computational Sensing and Robotics, Johns Hopkins University},
             addressline={},
             city={Baltimore},
             state={Maryland},
             postcode={21218},
             country={USA}}
             
\affiliation[hemi]{organization={Hopkins Extreme Materials Institute, Johns Hopkins University},
             addressline={},
             city={Baltimore},
             state={Maryland},
             postcode={21218},
             country={USA}}

%% Abstract
\begin{abstract}
Rapid developments in artificial intelligence and machine learning as applied to materials science are creating an urgent need for experimental data,  which can be provided by high-throughput and autonomous laboratories. To date most demonstrations of such laboratories have focused on functional materials, with less attention paid to structural materials. We present here the Artificial Intelligence in Materials Design Laboratory (\aimdl), an automated, high-throughput facility for characterizing the microstructure and properties of structural metals and ceramics, with an emphasis on materials in extreme environments. 

\aimdl has two custom instruments for characterization of structural materials: HELIX for shock studies of materials, and MAXIMA for \xray diffraction and \xray fluorescence spectroscopy. Specifically designed for high-throughput studies, HELIX and MAXIMA are each capable of collecting data at rates two to three orders of magnitude faster than conventional systems. A third experimental station, SPHINX, is a commercial nanoindenter modified for integration into the automated workflow of \aimdl. A user (which may be human or an AI agent) directs the experiments to be carried out by means of a centralized control program. The experimental stations are linked by a conveyance that moves samples around the lab, with a robot at each station for sample transfer in/out of the instrument. The experimental stations also communicate with a common data layer that streams data autonomously from each instrument to a data portal, where their arrival triggers automated workflows for data reduction and analysis. The processed data are immediately available to the human operator or agentic AI, forming a closed loop for rapid decision-making and experimental control.
\end{abstract}



%% Keywords
%\begin{keyword}



%\end{keyword}

\end{frontmatter}

%% Uncomment following line to enable line numbers
%\linenumbers

\input{1-Introduction.tex}
\input{2-Background.tex}
\input{3-layout.tex}
\input{4-Robotics.tex}
\input{5-MAXIMA.tex}

\input{6-HELIX.tex}

\input{7-Nanoindenter.tex}

\input{8-Data.tex}
\input{9-samples.tex}
\input{10-example.tex}

\input{11-discussion_and_conclusions.tex}

\input{12-endmatter.tex}

\bibliographystyle{elsarticle-num} 
\bibliography{aimdl.bib}

\end{document}

%% file: 1-Introduction.tex
\section{Introduction}
\label{sec:introduction}

Combinatorial and high-throughput methods for fabricating and characterizing materials have the potential to accelerate materials development by allowing rapid exploration of composition and processing space~\cite{GreenAPR:17,LudwigNPJCompMat:19,GregoireNS:23}. Recent  advances in machine learning (ML) and artificial intelligence (AI) have spurred the development of automated and autonomous materials research laboratories, which can provide capabilities for rapid iteration through AI/ML-guided design cycles as well as generation of the large datasets required for training models~\cite{HungDD:24,BayleyMatter:24,SzymanskiNature:23,PyzerKnappNPJCM:22,StachMatter:21}.

Most of these efforts have focused on functional materials. Although materials fabrication is never trivial, for functional materials the influence of microstructure on properties is limited and the fabrication processes themselves are often amenable to automation. In comparison, automated and autonomous research on structural materials has lagged, for three principal reasons: It is difficult to compute mechanical properties of interest; thermomechanical processing of structural materials and subsequent sample preparation is time-consuming and difficult to automate; and structural characterization is challenging~\cite{MiracleCOSSMS:24}. These issues are all related to the essential roles of microstructure in mechanical behavior. Microstructure is important on length scales ranging from nanometers to millimeters, necessitating multi-scale approaches to both computation and characterization. Furthermore, the size of a specimen can itself influence important behaviors such as strength, ductility, and toughness~\cite{GreerPMS:11,ArmstrongJMR:19,RitchiePhilTransA:23}.

In this article we describe the Artificial Intelligence in Materials Design Laboratory (\aimdl) at Johns Hopkins University, which we have built to address some of these challenges. \aimdl was conceived, designed, and constructed to enable automated (and ultimately autonomous) high-throughput characterization of bulk structural alloys and ceramics. \aimdl has two custom instruments specifically designed for this task: A laser-microflyer impact system for high-throughput shock studies at both ambient and elevated temperature, and a high-energy transmission \xray diffraction system for rapid assessment of bulk microstructure. It also incorporates automated nanoindentation for mapping quasi-static mechanical properties, and additional support stations (optical profilometry and laser etching). All of these capabilities are automated and physically integrated through a conveyance and robotics system that transports samples around the laboratory and hands them off to the individual instruments for measurements. Lab operations are controlled by a central run manager, which can receive its instructions from either human operators (via an application programming interface or a graphical web interface) or AI agents. A powerful and secure unified data streaming and processing architecture enables autonomous data analysis, flexible workflows, and automatic ingestion of processed data by AI/ML algorithms. Finally, we describe a processing capability to produce bulk combinatorial specimens of structural alloys, and demonstrate the capabilities of \aimdl for high-throughput characterization of such samples.

%% file: 2-background.tex
\section{Background}
\label{sec:background}
The focus of \aimdl is on mechanical properties and behavior of structural materials, especially under extreme conditions of strain rate, pressure, and temperature. This focus, combined with the goal of achieving high throughput for model training and efficient exploration of parameter space, influenced all of the key design choices for the laboratory.
\begin{figure*}[tb]
\centering
\includegraphics{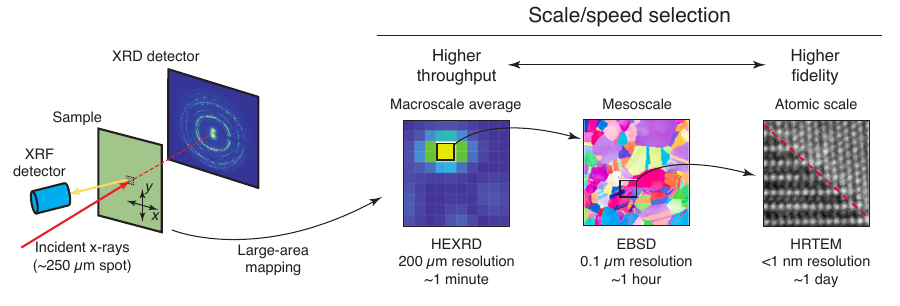}
\caption{Trading off speed and fidelity in microstructural characterization. High-energy x-ray diffraction (HEXRD) in \aimdl provides exceptionally high throughput (thousands of scans per day) but at relatively poor spatial resolution and with limited fidelity. However, it provides the ability to rapidly screen samples for regions of interest, which may then be examined at increasing levels of detail (but more slowly and at higher cost) with techniques such as electron backscatter diffraction (EBSD) and high-resolution transmission electron microscopy (HRTEM). (Sample EBSD data from Ref.~\cite{DellaVenturaUM:25} and HRTEM image from Ref.~\cite{XieSA:21}.)
\label{fig:fidelity_vs_speed}}
\end{figure*}

There are numerous examples of automated, autonomous, and high-throughput laboratories for chemistry and materials science. In autonomous labs where materials synthesis and processing is performed, the emphasis is often on using either solid- or liquid-state processing to produce crystalline compounds~\cite{SzymanskiNature:23,LuntCS:23,DaiNature:24,SongJACS:25}. In most such laboratories structural characterization is limited to identifying the phase(s) produced, usually by \xray diffraction. One reason for this is that the materials produced in these labs are of interest primarily for functional properties that are relatively insensitive to microstructure. Similar comments can be made with regard to the physical form of the materials produced in autonomous and high-throughput laboratories. If the primary consideration is which crystalline phases have been synthesized, the physical form of the product (powder or thin film, for example) and its characteristic dimensions (particle size or film thickness) are of limited importance. 

The situation is entirely different for mechanical properties and behavior, which are sensitive to both microstructure and specimen length scales. This is of course true for characterization, but it is also important to note its relevance for fabrication and processing of structural materials, especially metals. The microstructure of a vapor-deposited thin film or compacted powder is different from that of a metal produced by traditional thermomechanical (rolling, extrusion, forging, etc.) or additive manufacturing processes used to produce bulk structural metals. These considerations drove us to develop a laboratory capable of testing and characterizing samples of bulk structural materials.

Another prime consideration in developing \aimdl was achieving the highest possible throughput, to rapidly generate large amounts of data for materials exploration campaigns and training AI/ML models. As an example of how this emphasis drove key design decisions, consider that for microstructural characterization there is inherently a trade-off between throughput and fidelity, as illustrated in  Fig.~\ref{fig:fidelity_vs_speed}. The most detailed information is provided by techniques such as transmission electron microscopy (TEM) and scanning transmission electron microscopy (STEM) with atomic or near-atomic resolution. The high fidelity of these techniques comes at a price, most notably the small volume of material examined and the extensive sample preparation required, which limits their throughput and renders them unsuited to full automation. Electron backscatter diffraction (EBSD) examines larger volumes (albeit at poorer resolution) and is a workhorse technique for microstructural characterization of metals, but here too sample preparation is a concern. Although EBSD can be performed on bulk specimens, careful preparation of the surface is required. This is time consuming and problematic if many different types of samples are to be examined (each potentially requiring a different recipe for surface preparation  which may be unknown at the start of the experiment). In \aimdl we therefore took a different approach, knowingly sacrificing some of the rich detail available from EBSD in favor of a technique (transmission high-energy \xray diffraction) that requires little or no sample preparation and thus achieves dramatically higher throughput. 

Characterizing a large number of samples is of limited utility, however, if the data obtained are not available immediately in forms suitable for ingestion by AI/ML algorithms and decision-making (by either humans or AI agents). Laboratory information management systems (LIMS) that facilitate acquisition, analysis, and management of laboratory data have a long history~\cite{astm_e1578,SkobelevMT:11}. The challenges associated with these tasks are multiplied in autonomous high-throughput laboratories where a large number of samples must be individually identified, tracked, and moved among instruments automatically; complex, multi-step workflows are planned and executed; and decisions are made based on resulting data in real time. Careful consideration of these data flows is an essential component of planning an autonomous high-throughput laboratory.

%% file: 3-layout.tex
\section{\aimdl layout and general operation}
\label{sec:layout}
\begin{figure*}
\centering
\includegraphics[width=0.85\linewidth]{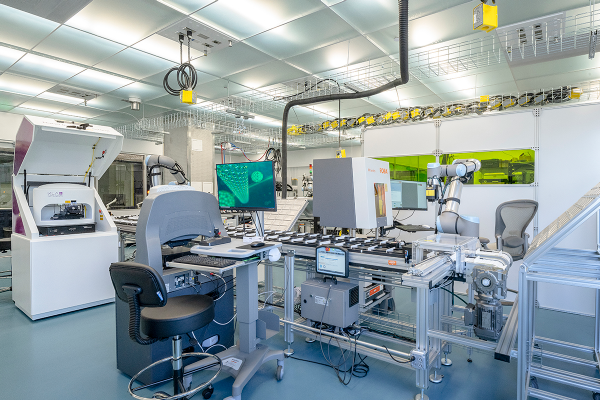}
\includegraphics[width=0.85\linewidth]{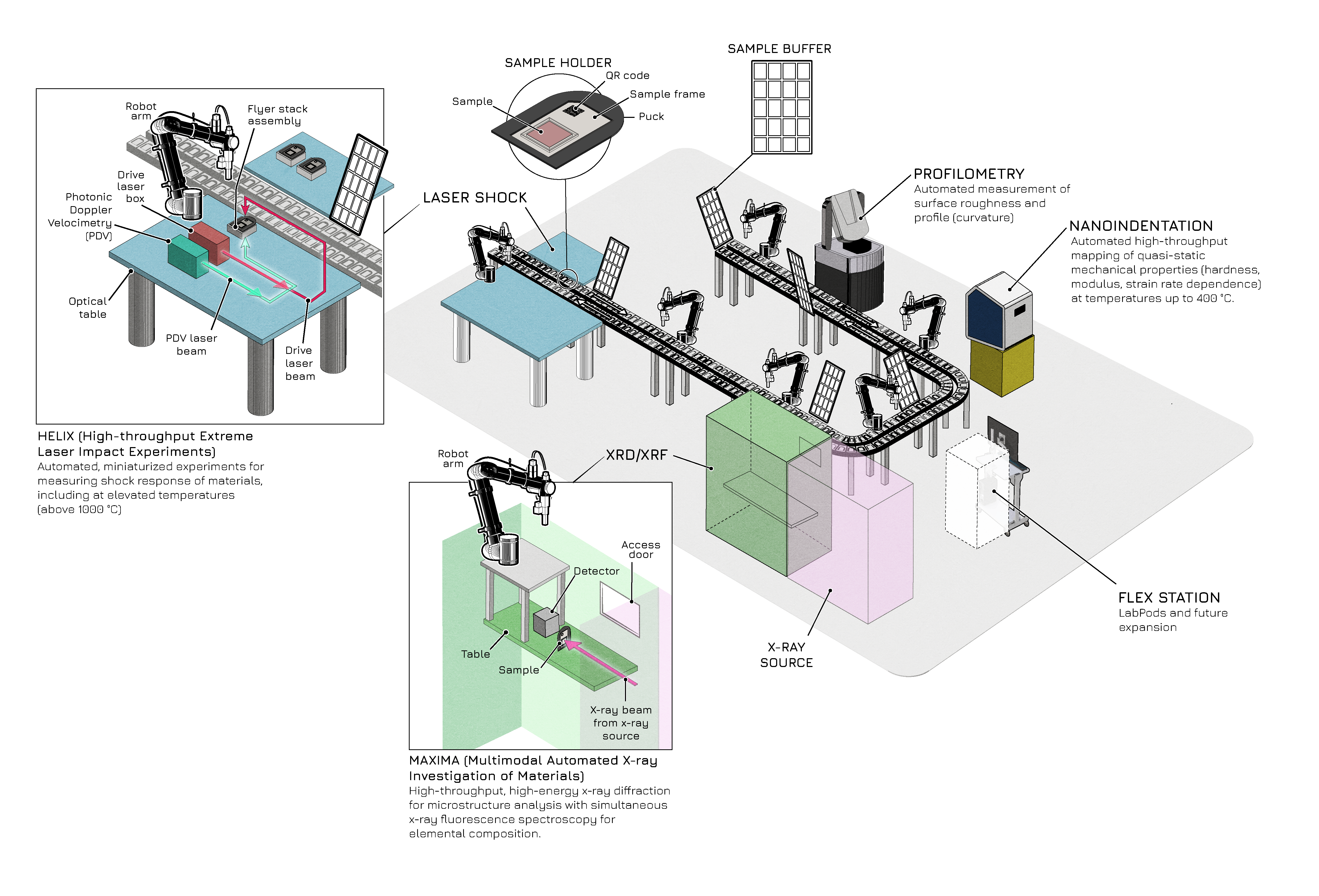}
\caption{The Artificial Intelligence for Materials Design Laboratory (\aimdl) at Johns Hopkins University comprises five experimental stations surrounding a central conveyance and robotic sample handling system (Sec.~\ref{sec:robotics}). The experimental stations include \mbox{\xray} diffraction and \mbox{\xray} fluorescence (MAXIMA, Sec.~\ref{sec:maxima}), laser microflyer impact for shock studies (HELIX, Sec.~\ref{sec:helix}), and nanoindentation (SPHINX, Sec.~\ref{sec:nanoindentation}). Profilometry is available for characterization of surface topography, and the flex station allows for incorporation of new capabilities.
\label{fig:aimdl_layout}}
\end{figure*}

The hardware of \aimdl consists of five stations surrounding a central sample handling and conveyance system (Fig.~\ref{fig:aimdl_layout}). The primary experimental stations (\mbox{\xray} characterization, laser shock, and nanoindentation) are described in detail in Sections~\ref{sec:maxima}-\ref{sec:nanoindentation} below. A fourth experimental station for optical profilometry is not described here. Finally, there is space reserved for flexible incorporation of instruments to provide new capabilities for specific experimental campaigns.

\begin{figure}
\centering
\includegraphics[width=0.95\columnwidth]{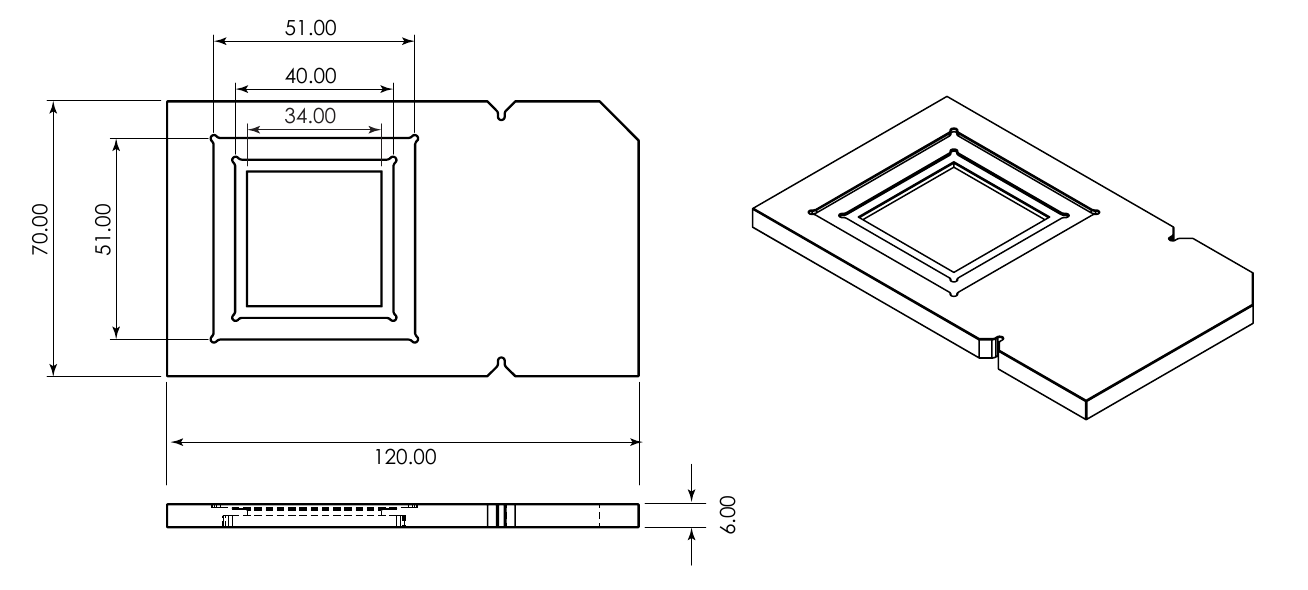}
\caption{Standard holder for samples in \aimdl. The holder has a $\qty{34}{mm}\times \qty{34}{mm}$ square through-hole, allowing access to both front and back surfaces for experiments. Samples are held securely by either adhesive or tape. The large flat area to the right of the square hole is to provide a space for the robotic vacuum grippers to pick up the holder.
\label{fig:sample_holder}}
\end{figure}

To facilitate robotic operation of the lab, samples are secured in a standard holder (Fig.~\ref{fig:sample_holder}). The sample holders are designed to provide access to both surfaces of the specimens, which is important for both the laser shock and \xray diffraction measurements. The typical sample form factor is a $\qty{40}{mm}\times\qty{40}{mm}$ foil, $\approx\qtyrange{100}{400}{\micro m}$ thick. The choice of thickness was motivated by two factors. First, we wanted to be able to examine samples with microstructures reflective of bulk thermophysical processing techniques (such as rolling and extrusion), as opposed to microstructures typical of vapor-deposited thin films. Second, we were aware that some mechanical behaviors of interest (spall failure due to shock, for example) are influenced by the length scale of the specimen. The lateral dimensions of the samples were chosen for operational convenience, but one motivating idea was that we could produce compositionally-graded specimens for combinatorial studies~\cite{GregoireNS:23} as discussed in more detail in Sec.~\ref{sec:example} below. 

Each sample entering the lab receives a globally unique persistent identifier, using the International Generic Sample Naming system~\cite{igsn} (if it does not already have one). Data produced by an experimental campaign are stored in a data portal, which can be hosted locally or remotely, and the IGSN for each sample is uniquely associated with an entry in the portal. Each sample holder has a QR code linking the physical specimen to the entry in the data portal (and thus to all of the data and metadata associated with it). The data architecture for \aimdl is described in more detail in Sec.~\ref{sec:data} below.

The movement of samples through the lab is coordinated by a central run manager (Sec.~\ref{sec:data}). Samples are introduced via an input buffer and are automatically routed through a shared conveyance system according to the active workflow schedule. As samples progress through the lab, the run manager tracks their state and coordinates transfers to and from experimental stations, including temporary staging when required to accommodate differences in experimental latency. The physical implementation of the robotic sample handling and conveyance system is described in detail in Sec.~\ref{sec:robotics}.

\begin{figure*}
\centering
\includegraphics[width=0.85\linewidth]{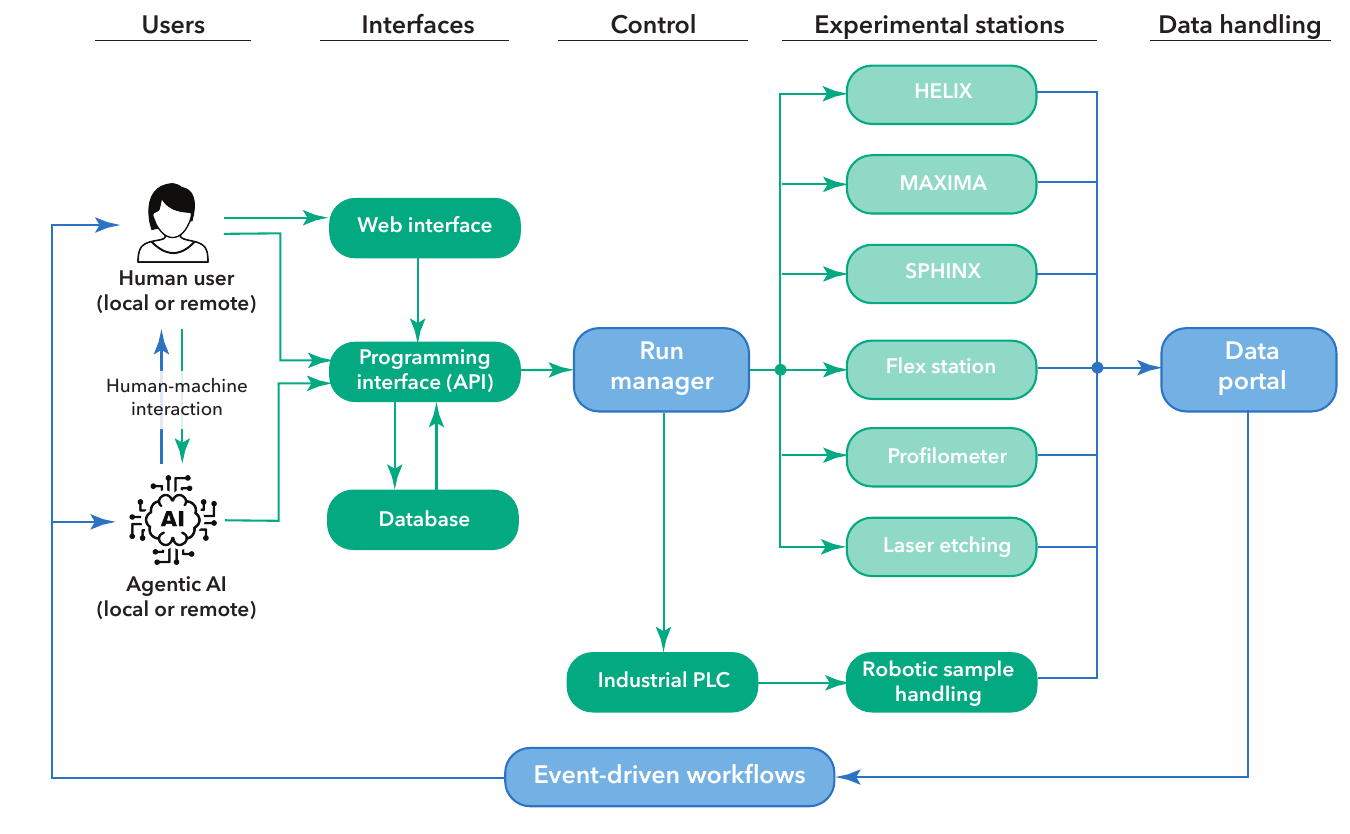}
\caption{Control signals and data flow in \aimdl. A human or AI agent provides instructions to the run manager through an API (either directly or, for the human, through a web interface). The run manager directs the flow of samples through the lab and issues instructions to the various instruments on the experiments to be run. Data and metadata are streamed autonomously off the instruments to a data portal, where their arrival can trigger event-driven workflows such as automated data processing. The data and metadata are immediately available to the agent (human or AI) to enable decisions to be made for the next cycle of experiments.
\label{fig:signals}}
\end{figure*}

The control signals and data flow in \aimdl are outlined in Fig.~\ref{fig:signals} and described in more detail in Sec.~\ref{sec:data}. The run manager receives its instructions from either a human user or an AI agent by means of an application programming interface (API). If desired, a human user can also interact with the run manager through a web interface built on the API. In addition to controlling the flow of samples through the lab, the run manager provides instructions to the various experimental stations on the tests to be run. Data and metadata are streamed autonomously from each instrument to a data portal, and can trigger automatic workflows such as data and metadata curation and processing. The data and metadata are immediately available to the human or AI agent directing the lab, allowing for closed-loop control.

%% file: 4-robotics.tex
\section{Sample handling and robotics}
\label{sec:robotics}
The physical handling of samples in \aimdl is managed through an integrated robotic and conveyance system. Samples are robotically transferred onto a conveyance consisting of a continuous belt that transports individual plastic pucks, each serving as a carrier for a single sample holder. The conveyance is capable of carrying up to 225 sample holders at one time, transporting each in sequence to the experimental stations. At each end of the loop, a transfer station equipped with a clamping actuator secures and aligns each puck for accurate robotic loading and unloading. A buffer tray at each station provides an intermediate staging area for samples during queued or multi-step operations.

%Once a sample is placed on the input buffer, the run manager initiates a robotic transfer of the corresponding sample holder onto the conveyance loop. The conveyance system can carry up to 225 samples at a time and moves them sequentially across the experimental stations according to the active workflow schedule. %Each sample holder carries a mounted 40 mm $\times$ 40 mm metallic foil, and the system maintains precise orientation and traceability throughout its movement across the lab. 

Six collaborative manipulators (UR10e, Universal Robots) perform all sample handling operations across \aimdl. Each robot is equipped with a vacuum gripper and a barcode reader for reading a QR code on the sample holder. The barcode system enables each robot to identify the sample holder, verify its assigned process, and log task completion to the central run manager. Four of the experimental stations use a single robot for sample loading and unloading. At the HELIX station, two robots are used: One for sample loading/unloading, and one for assembly and disassembly of the sample launch package (Sec.~\ref{sec:helix}). 

For sample loading and unloading at each station the robots execute coordinated pick-and-place motions, lifting sample holders from the pucks and delivering them either to the instrument or the associated buffer tray. A programmable logic controller (PLC) coordinates these robotic actions with the conveyance system and communicates task flags and sample identifiers to the central run manager to maintain synchronized operation across all stations.

%% file: 5-maxima.tex
\section{MAXIMA: Microstructural characterization}
\label{sec:maxima}

As discussed in Section~\ref{sec:background}, in microstructural characterization there is a tradeoff between fidelity (level of detail) and throughput. For researchers interested in mechanical behavior of materials, the richness of information provided by EBSD often makes it the tool of choice, but it has some limitations for high-throughput studies of bulk materials. Besides the need for surface preparation mentioned in Sec.~\ref{sec:background}, because EBSD is only sensitive to near-surface structure and scanning large areas is relatively slow, the information obtained may not be representative of the bulk microstructure. Furthermore, the large specimen tilt required for EBSD may limit studies of large combinatorial specimens, identification of unknown phases is still a research problem, and there is considerable overhead associated with the need to conduct the work in vacuum.

Our approach to microstructural characterization in \aimdl uses transmission high-energy \xray diffraction instead, which has numerous advantages for high-throughput characterization: Little or no sample preparation is required, bulk microstructures can be examined, phase identification for \xray diffraction is largely a solved problem, and large-format samples are easily scanned. The tradeoff relative to EBSD is that the microstructural information obtained is not as rich. However, once a region of interest has been identified, other, higher-fidelity characterization tools may be brought to bear.

\begin{figure}
\includegraphics[width=0.9\columnwidth]{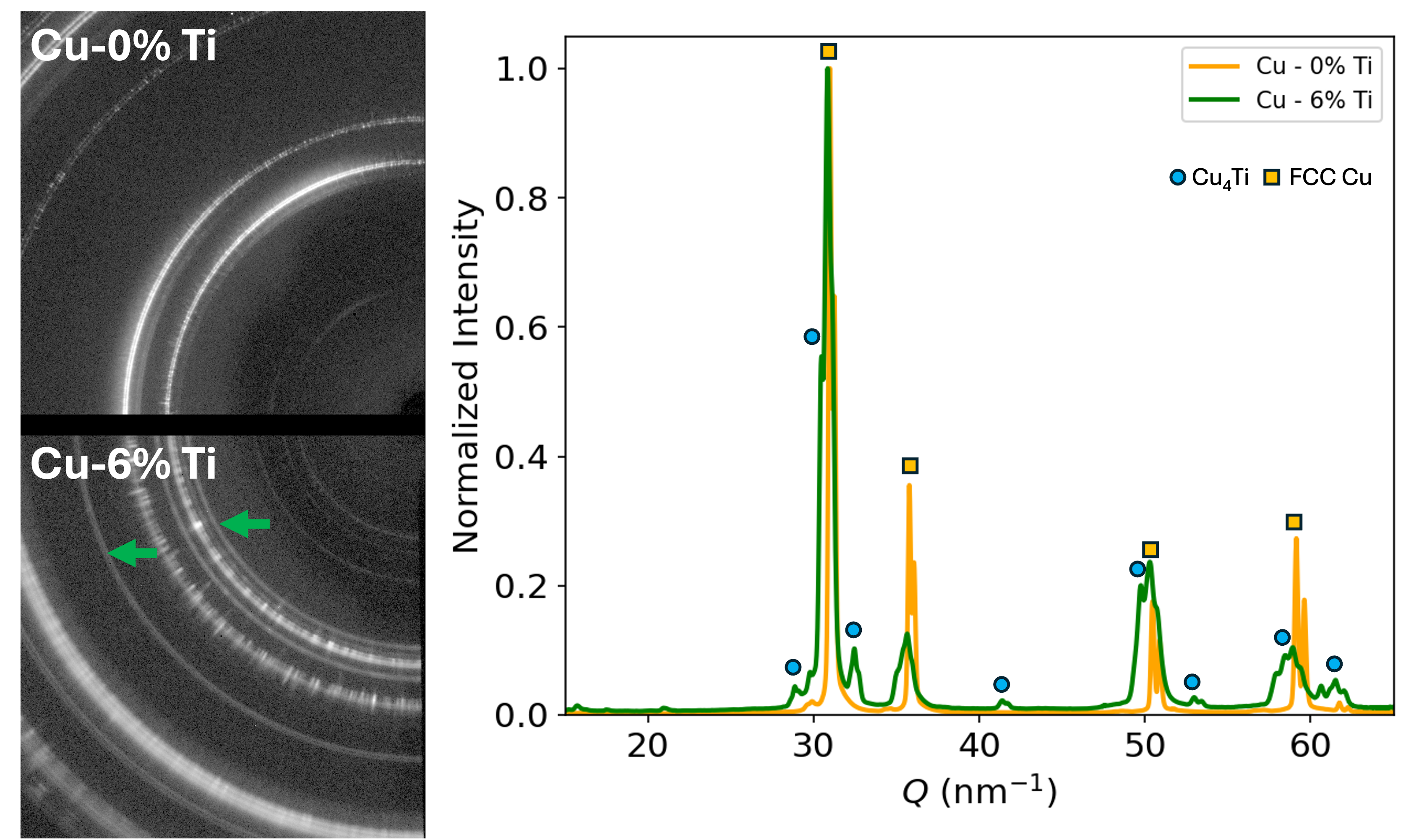}
    \caption{Sample XRD pattern portions showing the bottom and top of the combinatorial samples. Increasing Ti content introduces $\rm{Cu_4Ti}$ into the microstructure, which can be seen in the formation of faint rings (labeled by green arrows). \qty{100}{s} scans performed at \qty{700}{W}.}
    \label{fig:maxima}
\end{figure}

The instrument we designed is called MAXIMA (Multimodal Automated X-ray Investigation of Materials) and is described in detail elsewhere~\cite{ParkMaxima}. The basic concept (illustrated in Fig.~\ref{fig:maxima}) is to provide a focused, high-energy \xray beam and record the diffraction pattern in transmission through the specimen. The transmission geometry means that we are measuring the true bulk (through-thickness) microstructure, and allows us to achieve spatial resolution of \qty{250}{\micro m}. This is a convenient scale because it is large enough to ensure that the sampled volume is representative of the microstructure as a whole in most cases, while still being small enough to allow spatially-resolved studies of combinatorial specimens.

Transmission \xray diffraction measurements in MAXIMA are made using an intense, focused, high-energy (\qty{24}{keV}) beam, with diffraction patterns recorded on a pixel array detector. This combination of source and detector allows us to record actionable diffraction patterns (good enough for phase identification) in as little as \qty{1}{s}. In addition to diffraction, the bright \xray spot on the specimen produces \xray fluorescence (XRF), enabling spatially-resolved measurements of composition to be made simultaneously with the diffraction measurements. This is particularly useful for studies of combinatorial specimens (as demonstrated in Sec.~\ref{sec:example} below). However, because these experiments are conducted in air, XRF spectroscopy measurements cannot reliably be made for elements with atomic numbers below 12 (Mg). Furthermore, because the system geometry was optimized for XRD and not XRF, in experiments where both are measured the throughput is limited by the XRF count rate.

In a typical experiment a robot transfers a sample holder from a buffer tray (Fig.~\ref{fig:aimdl_layout}) into MAXIMA through an automatic safety door, placing it in a receptacle. After the door closes a UR3e robot inside MAXIMA picks up the sample holder and positions the specimen between the \xray source and detector for the measurement. The \xray shutter opens, and the robot rasters the sample perpendicular to the \xray beam as necessary to record data from different regions of the specimen. When the scans are complete the shutter closes and the internal robot returns the sample holder to the receptacle, where it is retrieved and returned to the buffer by the external robot.

Both the diffraction and fluorescence data, together with the associated metadata, are automatically streamed from the local MAXIMA computer to the data portal, using OpenMSIstream~\cite{EminizerJOSS:23} (Sec.~\ref{sec:data}). The arrival of the data at the portal triggers automated workflows which use prerecorded standard data to calibrate the diffraction pattern (for the sample-to-detector distance), correct for detector position and tilt, find the beam center, and integrate the 2D images of the diffraction patterns into 1D (intensity \emph{v.} scattering vector magnitude) plots that are more convenient for phase identification.

The microstructural information provided by MAXIMA includes the identities and relative amounts of crystalline phase(s) present in the scattering volume, together with the lattice parameter and some indication of crystallographic texture (which influences the distribution of intensity around the diffraction rings, Fig.~\ref{fig:maxima}). It may be possible to infer some idea of the grain size from the spottiness of the patterns, but grain size determination by the Williamson-Hall method~\cite{WilliamsonAM:53} is limited to nanocrystalline specimens.

As an example of the XRD data produced by MAXIMA, Fig.~\ref{fig:maxima} shows scans of two \qty{250}{\micro m} thick sputter-deposited foils, one nominally pure Cu and one Cu-\qty{7}{at.\% Ti}. The XRD patterns show that the introduction of Ti shifts the FCC Cu peaks to smaller scattering vector ($Q$) values and introduces a new phase ($\rm{Cu_4Ti}$). There is also peak broadening, suggestive of a decrease in grain size, lattice strain, or both.

%% file: 6-helix.tex
\section{HELIX: High-throughput impact experiments}
\label{sec:helix}
Common approaches to determining the dynamic behavior of materials include plate impact, Kolsky bar (split-Hopkinson pressure bar), and direct impact~\cite{RameshChapter}. Although these methods are well established and provide high-fidelity data, they are expensive and are generally low throughput, partly due to the need for a high level of expertise and direct human involvement. Thus, these methods do not lend themselves to rapid exploration of a materials/processing space or generation of the large amounts of data necessary to train AI/ML models.  

Our approach in \aimdl takes advantage of the fact that dynamic experiments are inherently fast; therefore, we can dramatically increase throughput by automating the testing process and data handling. The HELIX (High-throughput Extreme Laser Impact eXperi\-ments) station in \aimdl provides a unique capability for rapid and inexpensive exploration of the dynamic properties of a wide range of materials. Described in detail elsewhere~\cite{Bhattacharjee:2025,wanchoo2026helix}, HELIX is essentially a miniaturized version of a conventional (gas or powder gun) plate impact facility~\cite{RameshChapter}. In HELIX the flyer plates are thin, \qty{1.5}{\milli\metre} diameter circular disks driven by laser impulse to impact velocities of up to $\sim \qty{2000}{\metre\per\second}$. The flyer plates generate shock waves upon impact with the target material, enabling measurement of material response under extreme conditions. The value of the plate impact approach is that the deformations are inherently one-dimensional, allowing the stress and deformation states in the material to be determined accurately. In this sense we gain throughput without losing fidelity.

\begin{figure}[tb!]
    \centering
    \includegraphics[width=1\linewidth]{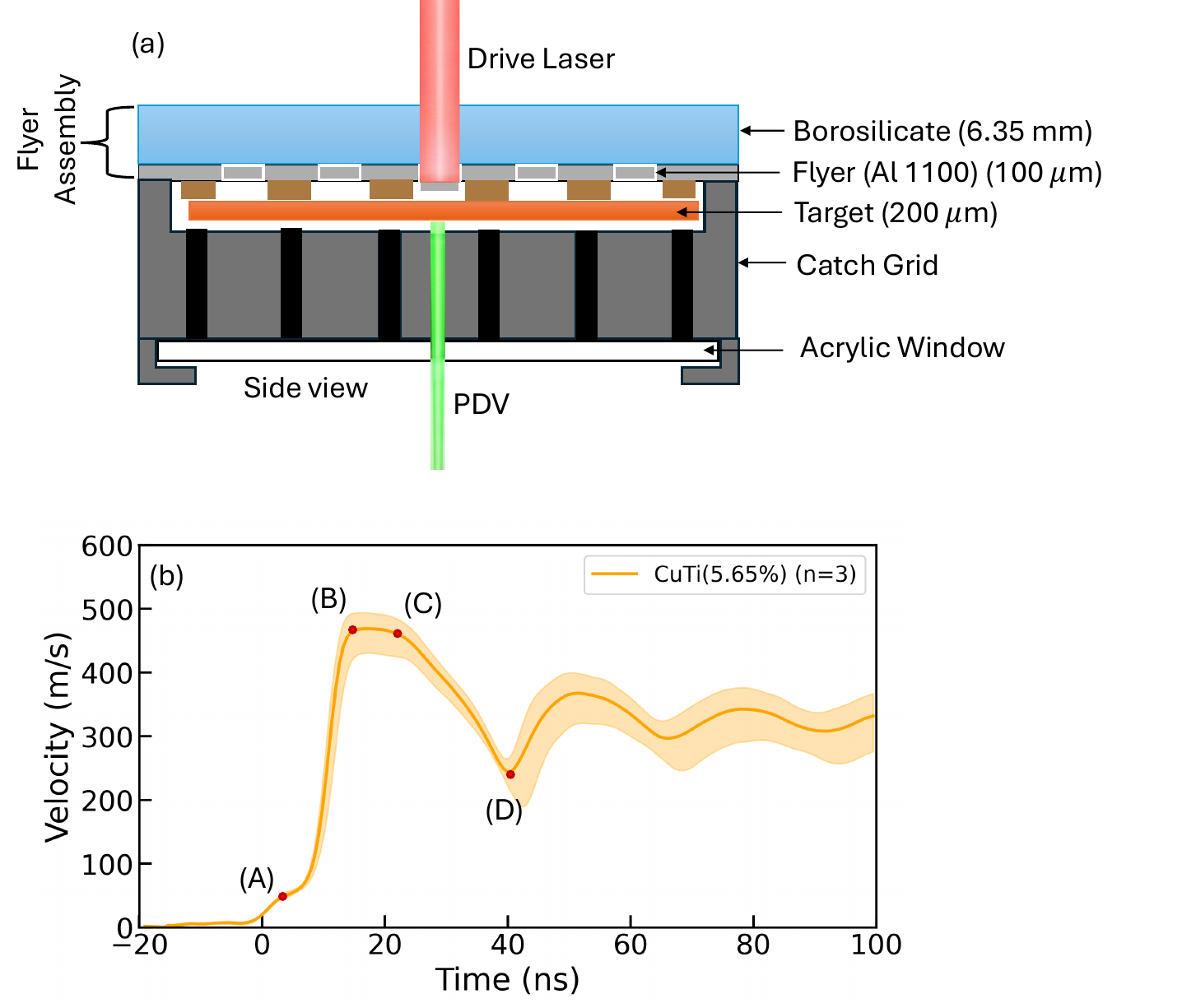}
    \caption{(a)~ Cross-section view of the launch package assembly in HELIX. The drive laser propels a \qty{1.5}{mm} precut laser flyer, which impacts the target material, while photon oppler velocimetry (PDV) is used to measure the free surface velocity of the target face opposite to the impact face. (b)~Free surface velocity trace obtained using PDV for a Cu-Ti sample (\qty{5.65}{at.\% Ti}). Point (A) marks the Hugoniot elastic limit (HEL), the region between (B) and (C) defines the shock stress, and the difference in free surface velocity between (C) and (D) is used to calculate the spall strength.}
    \label{fig:helix}
\end{figure}

The drive laser beam is incident on a `launch package', as shown in Fig.~\ref{fig:helix}(a). The launch package consists of two parts, separated by a spacer: a ``flyer stack" and the target material itself in the form of a thin sheet. The flyer stack is a thin sheet (\qtyrange{25}{200}{\micro\metre} thick) of the flyer plate material (typically aluminum), glued onto a glass plate with a uniform layer of epoxy. Each flyer stack is used to generate 25 flyer plates in a $5\times 5$ array, with the flyer disks predefined by laser cutting rings with a \qty{1.5}{mm} inner diameter into the flyer material sheet. Thus each test assembly allows us to run a total of 25 impact experiments on each target sample, after which the sample assembly is removed and a new one loaded. 

During an impact test (a `shot') the drive laser ablates the epoxy behind one of the predefined flyer disks, causing it to delaminate. This disk is the flyer plate, which then impacts the target at \qtyrange{100}{2000}{m.s^{-1}}, generating shock waves in the target (as well as in the flyer itself). The motion of the rear surface (opposite the impacted side) of the target plate is measured using PDV, providing the history of the particle velocity at the rear surface (Fig.~\ref{fig:helix}(b)). From the measured particle velocity at the rear surface we can use the standard analysis for plate impact experiments to determine such quantities as the Hugoniot elastic limit (HEL) and the spall strength of the target material~\cite{Wanchoo2026_HTSpall,wanchoo2026helix}. 

In addition to the rear surface velocity history, several other diagnostics are automatically measured and recorded for every shot, including the beam energy, beam profile, position on the target plate, and temperature of the sample. Metadata associated with the flyer stack and the sample are also recorded.  All of the data and metadata are automatically streamed and processed as described in Sec.~\ref{sec:data}. For spall measurements specifically, the data workflow includes automatic determination of the spall strength using custom, open-source software~\cite{Diamond_Ramesh_2024}.

% TCH 2026-02-19
% A few sentences deleted because they did not fit into the flow, but which we might want to add back:

%The key to the experiment is tuning of the drive laser beam and beam energy, precise triggering, and synchronization of the various systems involved in the experiment.

%The system enables high-throughput measurement of material properties and failure mechanisms under high pressures and temperatures and extreme strain rates (of the order of $10^6$ per second).

This process has been fully automated~\cite{Bhattacharjee:2025} to achieve truly high throughput. HELIX interacts with the \aimdl central conveyance system through two robots, as discussed in Sec.~\ref{sec:robotics}. The first robot picks up a sample holder from the conveyance and places it on the local buffer tray. The second robot then transfers the holder from the buffer tray to a table where it assembles the launch package consisting of the flyer stack, spacer, sample holder, and a recovery grid (which captures any debris released from the rear surface of the specimen as a result of the shock). The assembled launch package is then transferred by the robot to the optic table, where a high-precision stage moves the package to ensure that the appropriate point on the sample is being tested. Once the test is completed, the assembly process is reversed and the sample holder returned to the buffer and ultimately to the conveyance.

This high degree of automation of both the physical aspects of the experiment and the data streaming and analysis allows us to perform thousands of tests per day, compared to a handful of tests per day possible with conventional (gas or powder gun) experiments. This thousand-fold increase in throughput makes it possible to perform systematic discovery and design of materials. Further, the cost per test in HELIX is about 1000 times lower than the cost per conventional test.

%% file: 7-nanoindenter.tex
\section{SPHINX: Nanoindentation}
\label{sec:nanoindentation}
Nanoindentation plays an integral role in the \aimdl workflow, providing mechanical characterization of samples at loading rates lower than those of HELIX. It provides high spatial resolution measurements of hardness ($H$) and elastic modulus ($E$) from load-displacement curves, while continuous stiffness measurement (CSM) enables depth-dependent characterization~\cite{Oliver1992,SudharshanPhani2021MeasurementMeasurement,Pelletier2006CharacterizationTest}. These capabilities allow for rapid, localized evaluation of mechanical properties across graded materials.

%, where composition, microstructure, and properties vary spatially across a single sample. 

\begin{figure}[!t]
    \centering
   \includegraphics[width=\linewidth]{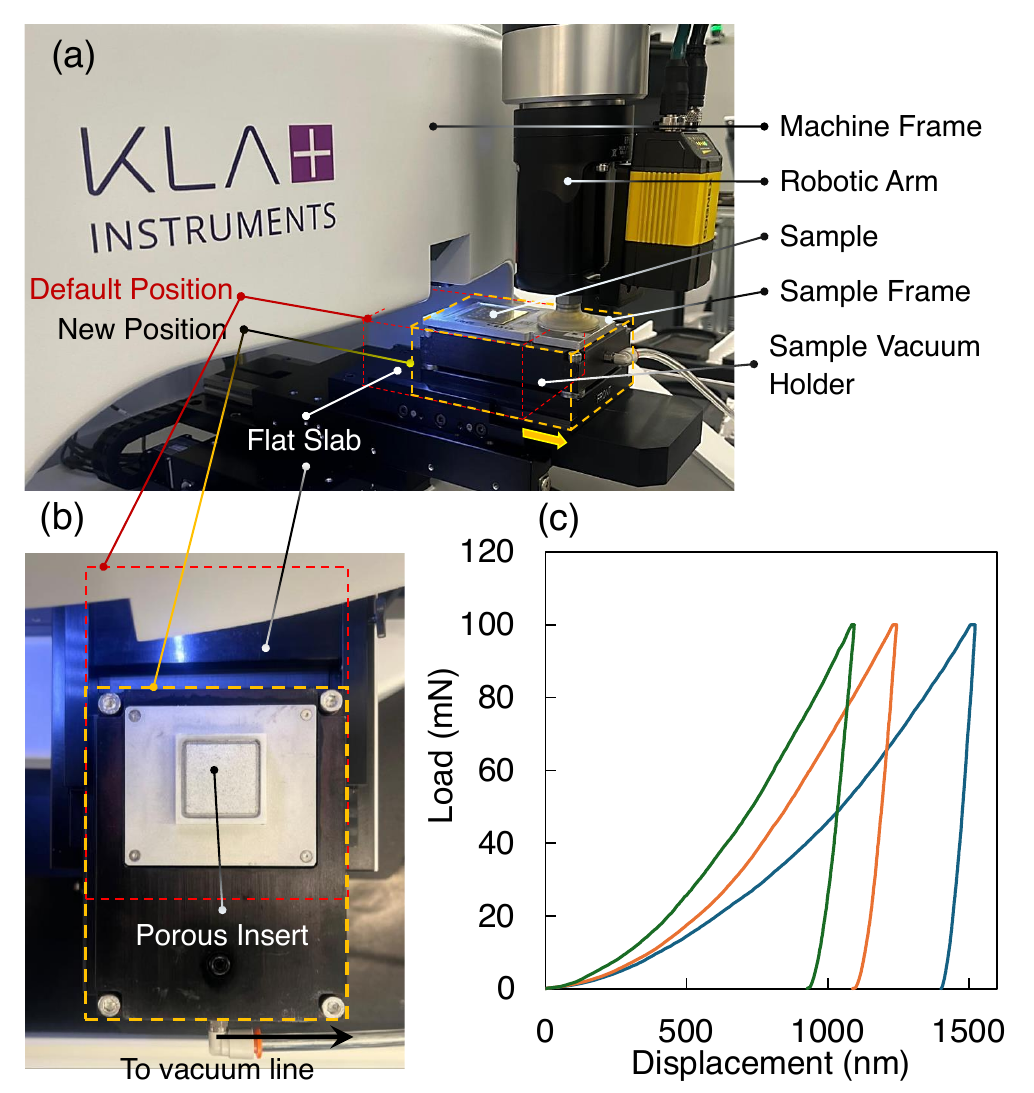}
    \caption{(a) The SPHINX system during robotic sample handling. The red and orange dashed lines highlight the default and new positions of the sample holder on the $x-y$ stage, respectively. The orange arrow indicates the direction of the displacement to allow the robotic arm to interface with the stage. (b) A top view of the sample vacuum holder. (c) Representative load–displacement curves from nanoindentation tests at different spots on a combinatorial Cu–Ti sample (with Ti content of 2–3 at.\% Ti) at a constant loading rate of \qty{10}{mN.s^{-1}} to a maximum load of \qty{100}{mN}.
}
    \label{fig:nanoindentation}
\end{figure}

Although commercial nanoindentation systems offer fast testing modes, imaging, and partial automation, several limitations hinder their use in fully autonomous, high-throughput workflows~\cite{Besharatloo2021InfluenceAlloys, Rossi2025,Chawla2025AutomatingAccuracy}. In particular, proprietary software ecosystems without user application programming interfaces (APIs) limit customization and integration with laboratory automation, while manual sample handling restricts scalability and reproducibility. To address these limitations, we developed a nanoindentation capability for \aimdl that we call Scanning Probe for High-resolution INdentation eXperiments, or SPHINX (Fig.~\ref{fig:nanoindentation}). SPHINX is based on a standard KLA G200X Nanoindenter with hardware and software modifications (implemented in collaboration with KLA) to enable integration with \aimdl operations for closed-loop automation. The standard G200X configuration includes low load (\qty{50}{mN}) and high load (\qty{10}{N}) actuators, each with a diamond Berkovich tip for measurements at ambient temperature. Advanced measurement modes enable rapid, high-density indentation mapping, while modular software supports CSM, strain-controlled, and load-controlled testing at strain rates ranging from \qty{1e-3}{\s^{-1}} to \qty{1e4}{s^{-1}}. Mechanical property mapping can be performed from room temperature up to \qty{500}{\celsius} using indenters with sapphire tips.

The standard G200X sample tray position does not provide sufficient clearance for robotic access to load and unload samples. To address this, we implemented hardware modifications on the dovetail rails of the $x-y$ stage by adding a flat slab to push the stage outside (Fig. \ref{fig:nanoindentation}(a)), repositioning the sample holder so that it is in the reachable workspace of the robotic arm. Complementary software modifications at the API level enabled direct control of stage coordinates, redefining sample loading and unloading positions to support autonomous operation.

In the standard \aimdl sample holder (Fig.~\ref{fig:sample_holder}), the sample is supported only around the edges to allow access to both sides of the sample as required for both MAXIMA and HELIX. However, because nanoindentation requires the sample to be rigidly supported from the back side, a custom sample stage for the nanoindenter was designed. As shown in Fig.~\ref{fig:nanoindentation}, the redesigned stage incorporates a porous aluminum foam insert. The sample with its holder is placed on the insert, and a vacuum is drawn from behind. This provides the necessary support for the sample during nanoindentation, while still allowing robotic loading/unloading and maintaining accessibility of both sides of the specimen at the other stations.

To enable integration with the \aimdl control scheme, a custom API suite for remote control of the nanoindenter was developed in collaboration with KLA. An Open Platform Communications Unified Architecture (OPC UA) client running on the nanoindenter control computer receives input parameters from the station namespace. These are passed to the station control code (Sec.~\ref{sec:data}) which communicates with the instrument. This architecture allows test parameters to be set without altering the instrument user interface, enabling remote operation and integration with AI-driven agents. At the end of each test the results are immediately available for review: they are written on disk, streamed off the instrument, and published with metadata to the data portal for lab-wide access and ML-ready provenance, as detailed in Sec.~\ref{sec:data}. Moreover, the automated analysis tightens the test loop by filtering failed tests and rejecting data points outside predefined acceptance thresholds prior to downstream processing, thereby reducing human-induced variability.
%\begin{displaymath}
%\begin{array}{c}
%\text{Sample loading by robot} \\
%\downarrow \\
%\text{Microscope check} \\
%\text{(if needed)} \\
%\downarrow \\
%\text{Indenter movement with calibrated offsets} \\
%\downarrow \\
%\text{Multisample staging} \\
%\text{(switch mode, new multisample, add sample, set input)} \\
%\downarrow \\
%\text{Start test} \\
%\downarrow \\
%\text{Status polling} \\
%\text{(InView status and location)}\\
%\downarrow \\
%\text{Sample unloading by robot}
%\end{array}
%\end{displaymath}

%As load-depth data are collected, the controller can adaptively stop indentation at a location once curves meet a confidence/variance target (avoiding oversampling), skip redundant repeats when overlays are equivalent, and keep re-runs aligned on identical coordinates. Each run captures the full context, including method files, inputs, calibrated offsets, precise coordinates, and raw outputs; 

One challenge with thick-foil samples processed by sputter deposition (Sec.~\ref{sec:samples}) is that residual stresses can induce substantial curvature, which the vacuum clamping system in SPHINX cannot overcome. For such a case, a flat foil is laser-cut and then bonded to a \qty{500}{\micro m} thick fused silica substrate before loading it onto SPHINX. This backing provides mechanical support for the sample during nanoindentation testing, improving measurement reliability and ensuring reliable sealing against the vacuum chuck. Fig.~\ref{fig:nanoindentation}(c) presents representative load–displacement curves for Cu–Ti samples with different Ti concentrations (2–3 at.\% Ti) recorded in this way.

Finally, we note that although the G200X supports high-temperature nanoindentation, this capability has not yet been automated. This will require a substantial redesign of the heating and temperature control architecture and is planned as a future extension of the SPHINX platform.

%% file: 8-data.tex
\section{Software, data, and integration}
\label{sec:data}

The ultimate goal of any laboratory is the creation of new insights from the data it generates. A central challenge in heterogeneous, high-throughput laboratories such as \aimdl is the automation of experimental workflows, including integrated data handling and semantic unification of high-volume, heterogeneous data generated by multiple instruments and measurement techniques. The linked datasets thus created are not only the source of new knowledge; they are also the foundation for the AI workflows that enable autonomous control of the laboratory itself. Thus, a fundamental challenge for any autonomous lab is integration of data and automation systems.
\begin{figure*}[tb!]
\centering
\includegraphics[width=0.9\linewidth]{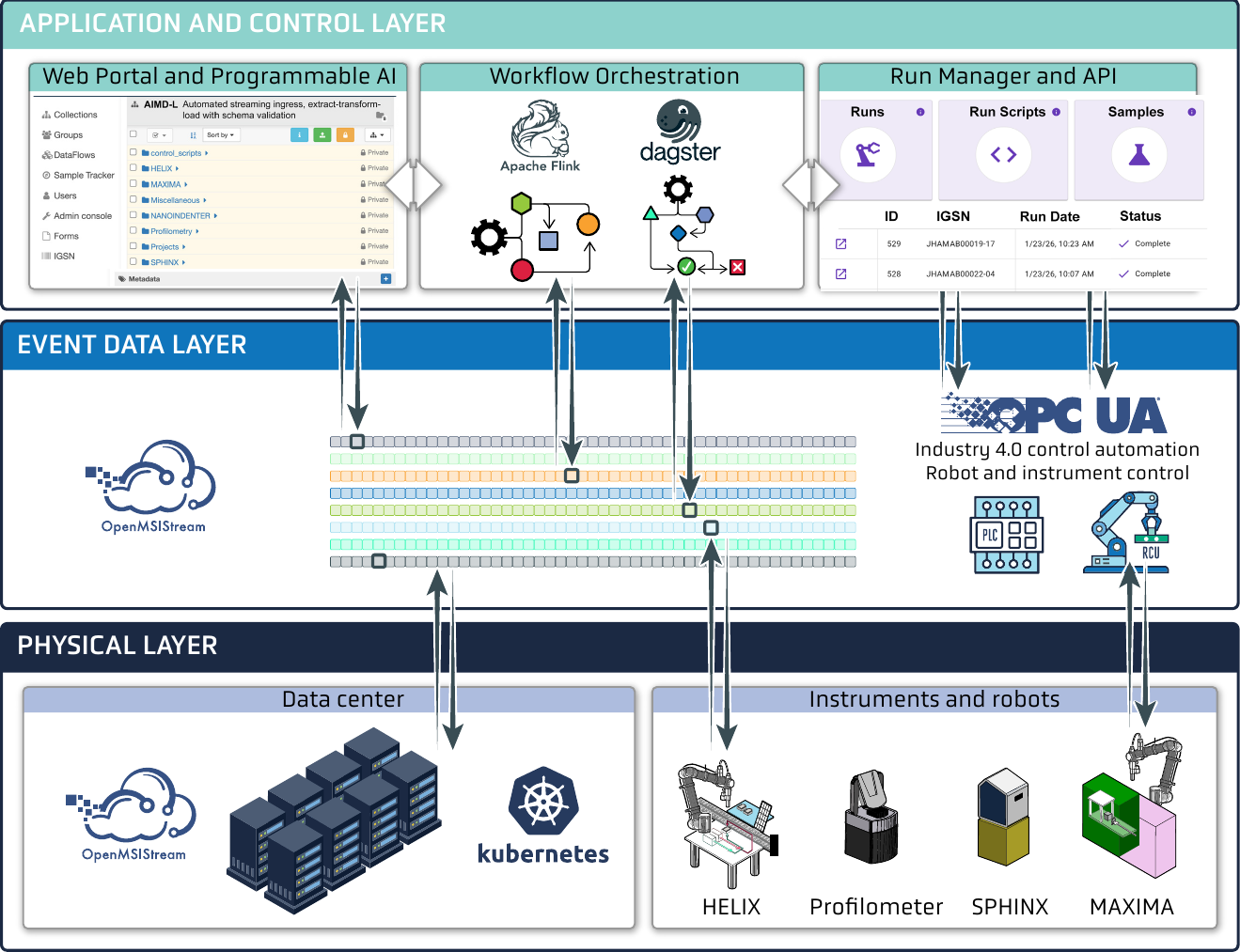}
\caption{Schematic \aimdl architecture (Sec.~\ref{sec:data}) highlighting a central event-driven, streaming data layer (middle) with data automation mediated via OpenMSIStream and automation of instruments and sample handling managed by the \aimdl server-client system integrated with OPC UA. This combination of event-based systems links the three components of the Application and Control Layer (top) and the data center and laboratory resources in the Physical Layer (bottom).}
\label{fig:aimdl_data}
\end{figure*}

\aimdl addresses this need with the architecture shown in Fig.~\ref{fig:aimdl_data}. The foundation is the physical layer---the data center, instruments, and robotics resources. This physical layer is linked to the top-level application and control layer through the event data layer. The event layer uses an OpenMSIStream~\cite{EminizerJOSS:23} event-driven backbone integrated with the \aimdl control software services and an Open Platform Communications Unified Architecture (OPC UA) server. 

A central design tenet in \aimdl is uniform programmatic control for both human and agentic users. User interaction from the application and control layer can be either programmatic (through a comprehensive API) or via a web-based interface (application service).  In this way, \aimdl allows teams to design, run, and analyze complex experiments through secure network access and provides the flexibility to accommodate both human and agentic users as peers. The data stack, which is built entirely from open-source solutions, enables users to interact with project data and computational workflows through a data portal~\cite{htmdec_deploy}. Workflows are implemented as real-time stream processing or orchestrated directed acyclic graphs.

 Among the challenges in engineering the \aimdl software and data system is integration and automation of diverse hardware and instrumentation developed by various equipment manufacturers with a diversity of network protocols, operating systems, and control abstractions. To ensure reliable and reproducible operation, the software of the \aimdl event data layer implements standardized communication mechanisms, coordinated sequencing of actions, and centralized tracking of system state. 

 The control software architecture of \aimdl consists of four coordinated services:
 \begin{enumerate}
 \item User interaction is provided through a web-based graphical interface (the \emph{application service}). 
 \item Programmatic access is exposed via an application programming interface (the \emph{data service} API), which authenticates users, validates experiment configurations and control definitions, and persists control metadata. 
 \item Execution and orchestration of experiments are handled by a controls server (the \emph{run manager}) based on OPC UA,  implemented in Python using the \texttt{asyncua} library. 
 \item Instrument-level control is provided by a collection of generalized Python OPC UA clients (the \emph{run clients}), each deployed in a Docker container on a dedicated station computer. 
 \end{enumerate}
Finally, a relational database management system stores control configurations, metadata, and execution history.

Laboratory operations are defined by \emph{run scripts}, which are structured, JSON-encoded instructional specifications exchanged through the data service. A run script identifies the station(s) to be employed and describes station-specific input parameters (such as scan coordinates and instrument settings). These scripts do not contain any sample-specific information and are, therefore, reusable on multiple sets of samples. Before being executed, the run script is validated and stored in a database as a permanent record. The run script specifications were designed to facilitate dynamic generation and adaptation by AI agents in the control layer. 

To track lab operations in \aimdl we define a \emph{run} as an execution of a run script on a specific set of physical samples. The run manager automates \aimdl operations by ingesting run information from users via the application or data service, parsing the instructions, and instantiating the corresponding control state within the OPC UA server. Widely used in industrial automation, OPC UA provides a stateful, hierarchical address space and shared control plane for communication among system components~\cite{OPCReference}. Within the run manager, this address space includes a system-level namespace that tracks runs, samples, and execution progress, as well as individual namespaces for each station that record assigned samples, input parameters, and operational status. State changes are propagated to clients through OPC UA subscriptions and monitored items, enabling structured, real-time feedback and coordination across distributed components as runs and stations transition through their operational states. By representing runs, stations, samples, and robotic state within a single OPC UA address space, the system provides a unified, inspectable model that supports real-time monitoring, fault recovery, and extensibility to new instrumentation.

In practice,  experimental control utilizes containerized run clients at each station.  Run clients receive OPC UA commands that invoke station-specific control logic while presenting a uniform interface. The underlying functionality and implementation varies by station. These are realized using instrument-specific software development kits, application programming interfaces, or direct socket-based communication. Robotics and the sample conveyance (Sec.~\ref{sec:robotics}) are managed by a programmable logic controller (PLC), which also tracks sample position throughout the system. The run manager communicates with the PLC to coordinate sample routing and to synchronize station ingress and egress.

During execution, the run manager assigns samples to stations and coordinates robotic delivery. Samples are transferred between the conveyance, buffer locations, and stations robotically (Sec.~\ref{sec:robotics}). When an experiment is completed, the sample is removed from the station, and experimental data are streamed (using OpenMSIStream~\cite{EminizerJOSS:23}) from instrument storage to autonomous processing and curation. The run manager employs an event-driven, asynchronous execution model that enables multiple runs and station workflows to progress concurrently while preserving deterministic sequencing of control actions. Explicit assignment priority queues are maintained for each station to manage sample execution order, optimizing system scheduling and throughput. Users can monitor run progress and overall system status in real time via feedback provided through the application service.

Scientific data handling and management in the data layer leverages OpenMSIStream~\cite{EminizerJOSS:23} tools for streaming data production at instrument stations (Fig.~\ref{fig:aimdl_data}). This event-driven paradigm provides high-speed (\qty{200}{Gb.s^{-1}}) asynchronous secure streaming to data topics in the Johns Hopkins Bloomberg Data Center, with \qty{1.5}{PB} of object storage and an additional \qty{400}{TB} of traditional filesystem storage. Apache Kafka data brokers in the data center provide dynamic scaling and topic replication for speed and reliability. Data consumption, stream processing, and autonomous curation in the data center are managed by OpenMSIStream and Dagster-orchestrated workflows running on two Kubernetes clusters.

User projects in \aimdl originate in the web portal with user group access/authorization and pluggable access to computational and dataflow tools. Unifying semantics use DOIs assigned to both projects and samples (Datacite IGSN DOIs \cite{igsn_manuscript}) to create a linked data model. Every sample that enters \aimdl either has a preexisting DataCite IGSN PID or is provided one by the data system. DataCite RelatedIdentifiers~\cite{datacite_relatedidentifiers} are used to encode metadata knowledge graphs and all related event data accessible through the public DataCite API~\cite{dataciteAPI}. By isolating access-controlled metadata, the knowledge graphs remain private until released by users or in compliance with funding agency public access plans. Our development of a linked data model with the public DOI system integrates scientific knowledge representation with data governance, providing \aimdl with full data provenance across sample histories and data lifecycle. We note that this architecture is readily extensible to include work across multiple, interconnected, programmable laboratories.

%% file: 9-samples.tex
\begin{figure*}[tb]
    \centering
    \includegraphics[scale=1]{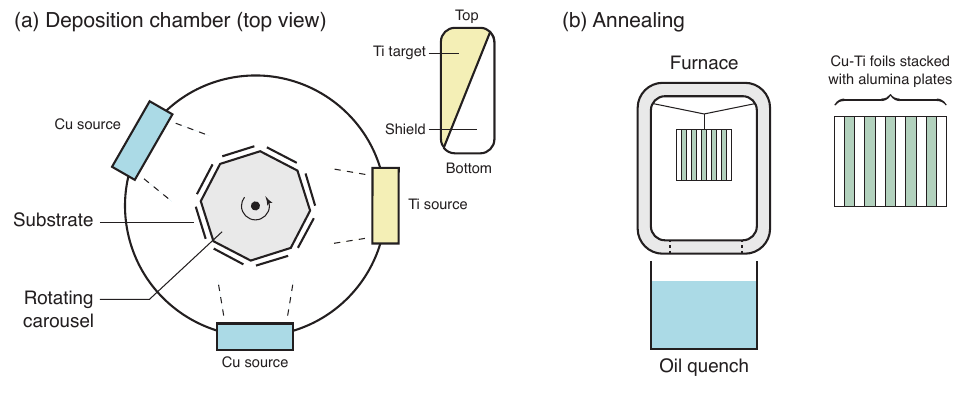}
    \caption{Overview of fabrication of compositionally-graded Cu--Ti alloy foils. (a) Sputter deposition of Cu--Ti foils onto brass substrates mounted on a carousel that rotated them past Cu and Ti sources in turn, producing nanolaminate specimens. The inset shows the shield on the Ti target, which causes a vertical concentration gradient from \qty{0.1}{at.\% Ti} at the bottom of the substrates to \qty{7}{at.\% Ti} at the top. (b) Following deposition the samples were removed from the brass substrates, sandwiched between alumina plates, placed in a holder, and suspended by a Nichrome wire for annealing (solutionizing) in a vacuum furnace. When annealing was complete the package was dropped directly into an oil quench bath.}
    \label{fig:sample_prep}
\end{figure*}

\section{Samples for combinatorial experiments}
\label{sec:samples}
The combination of the high-throughput characterization stations with the automatic conveyance, robotics, and data systems enables \aimdl to perform thousands of tests per day, on hundreds of specimens. To maximize the utility of this unique capability, it is advantageous not only to prepare a large number of specimens but also to ensure that the specimens span a wide range of chemistries and microstructures. This broadens the search space for materials discovery and expands the training domain for AI/ML models.

The sample form factor for \aimdl (Sec.~\ref{sec:layout}) was chosen in part to allow the characterization of combinatorial specimens, with gradients in composition, microstructure, or both. However, synthesizing such specimens at the scale required for \aimdl is a considerable challenge, especially considering that one of the objectives of \aimdl is to study specimens with microstructures and length scales representative of bulk (as opposed to thin film) materials. Here we demonstrate an experimental campaign in which we use large-scale magnetron sputtering to produce free-standing combinatorial metallic foils with thicknesses on the order of several hundred microns~\cite{BerliaMSEA:25}.

To obtain the results presented in the next section we sputter deposited compositionally graded Cu--Ti foils with Ti content varying from \qtyrange{0.1}{6.5}{at.\%} (Fig.~\ref{fig:sample_prep}). The foils were deposited onto large ($\qty{305}{mm}\times\qty{180}{mm}$) polished brass substrates, which were patterned with Kapton tape to define individual $\qty{40}{mm}\times\qty{40}{mm}$ samples (to follow the standard \aimdl sample form factor mentioned above). The substrates were mounted on a carousel that rotated past each of three sputtering targets in turn, two copper and one titanium, producing a multilayer structure with nanometer-scale layer thicknesses~\cite{BerliaMaterialia:26}. During deposition the Ti target was partially shielded to create a vertical composition gradient across each substrate. The total thickness of the foils was approximately \qty{200}{\micro m}, varying slightly across the foils because the shielding of the Ti target changes the overall deposition rate. Following deposition the specimens were removed from their substrates and solutionized by annealing under vacuum at \qty{900}{\celsius} for \qty{10}{hr}, followed by quenching into oil. This heat treatment both homogenizes the structure (eliminating the nanoscale layering from deposition) and coarsens the grain structure.

%% file: 10-example.tex
\section{Example data from AIMD-L}
\label{sec:example}
\begin{figure*}
\centering
    \includegraphics[width=0.75\textwidth]{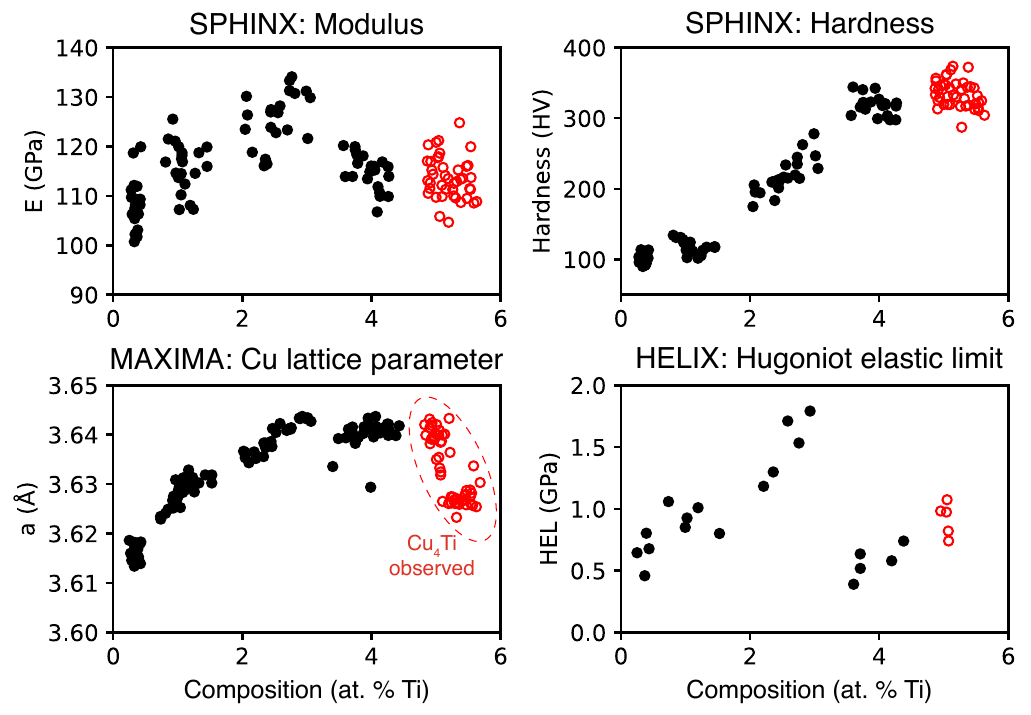}
    \caption{Example data from the three primary \aimdl experimental stations from combinatorial Cu-Ti samples: elastic modulus and hardness from nanoindentation measurements on SPHINX, lattice parameter of Cu from MAXIMA, and Hugoniot elastic limit (HEL) from HELIX. The results are plotted as functions of composition measured via XRF in MAXIMA. Red open circles indicate regions where where is clear evidence for the presence of Cu$_4$Ti intermetallic precipitates from the XRD measurements; black filled circles indicate regions without clear evidence for precipitates.}
    \label{fig:sample_data}
\end{figure*}

As a demonstration of the capabilities of \aimdl, we provide here example data obtained using the three primary experimental stations on large-scale combinatorial samples of Cu-Ti alloys of the kind described on the previous section. The example data shown in Fig.~\ref{fig:sample_data} include elastic behavior (modulus, from SPHINX), plastic behavior under low loading rates (hardness, also from SPHINX), onset of plasticity under shock loading (Hugoniot elastic limit, from HELIX), and the lattice parameter of fcc Cu solid solution (from MAXIMA). The data are all plotted as functions of composition, measured via XRF in MAXIMA. Also indicated on the plots are regions where there is clear evidence for the presence of Cu$_4$Ti precipitates from XRD in MAXIMA (though we note that there are limits on both the volume fraction and particle size at which precipitates can be identified using XRD). 

Several clear correlations emerge from these data. Modulus, HEL, and lattice parameter each show a maximum at around 3~at.\%~Ti, while hardness increases monotonically with Ti content over the range of compositions examined. It is particularly interesting that hardness and HEL, which are both measures of resistance to plastic deformation, show different trends. Although both are measured in compression, the strain rate and the strain state are significantly different between the two tests. Furthermore, the two methods sample the microstructure differently, with the nanoindentation measurements providing much more local response. Having such related yet distinct probes of mechanical response is likely to be useful when training AI/ML models from the datasets generated in \aimdl.

%% file: 11-discussion_and_conclusions.tex
\section{Discussion}
\label{sec:discussion}

Building an autonomous laboratory such as \aimdl presents challenges that extend well beyond configuring, integrating, and operating the individual instruments. For example, the supporting infrastructure required to house and interconnect multiple stations in a single facility demands careful, early-stage planning. These considerations include:
\begin{itemize}
\item Space: Moving samples around the lab requires additional floor space beyond the requirements of individual instruments. In particular, it is essential to plan for safe and efficient movement of humans without interfering with the robots during lab operations.
\item Cooling capacity: Multiple instruments in a confined space can produce significant thermal loads that exceed the capacity of standard laboratory air conditioning systems.
\item Air handling: Lasers and other precision equipment may require a low-particulate environment, making high-efficiency air filtration and flow control critical.
\item Electrical requirements: Each instrument will have its own voltage, current, and grounding requirements. To minimize electrical noise, separate circuits and grounds may need to be provided for different purposes. Data cables may need to be segregated from power distribution. Electrical distribution and cable management must be planned holistically to ensure safety, accessibility, reliability, and scalability.
\item Networking and data integration: Manufacturers may enforce specific IP ranges and virtual local networks for their instruments; integrating these may require bridging these networks while maintaining compliance with institutional and sponsor cybersecurity policies.
\item Safety: A laboratory with multiple instruments operating simultaneously multiplies the safety hazards, particularly since automated robotic operation may prevent the imposition of access limitations (often a key component of safety). Comprehensive interlock systems and automated signage linked to active equipment states may be required. Lab personnel must be trained to recognize a wide variety of hazards and in the appropriate responses, including emergency shutdown protocols.
\end{itemize}
Although it is impossible to anticipate every future need, proactive planning for these key infrastructure components can greatly reduce time, cost, and disruption once the laboratory is operational.

Because the central sample handling and robotics were designed to be flexible, \aimdl is capable of integrating new instruments through a ``flex'' station (Fig.~\ref{fig:aimdl_layout}) which provides access to the conveyance and a robot for sample handling. This allows us to expand and enhance the capabilities of \aimdl, limited only by conformance to the basic form factor of the sample holder and specimens. One example in the context of materials in extreme environments would be thermal measurements (conductivity and heat capacity) by laser flash techniques. Other possibilities include enhanced structural characterization (robotic sample polishing + optical or electron microscopy) and corrosion or oxidation resistance.

One limitation of \aimdl in its current configuration is that there is no way to manipulate the microstructure of the materials studied, and thus no ability for fully automated, closed-loop optimization studies. Simple thermal processing, such as spot laser annealing using a heating laser similar to that already incorporated into HELIX (Sec.~\ref{sec:helix} is already under consideration. But integrating more complex thermomechanical processing such as rolling or forging into \aimdl is challenging, particularly with regard to automated production of suitable specimens from bulk workpieces. Although the relatively simple form factor of the \aimdl specimens and holder (Fig.~\ref{fig:sample_holder}) makes it possible to envision solutions to these challenges, the level of effort varies with the characterization to be performed. For example, MAXIMA is relatively insensitive to sample preparation, and so one could imagine XRD studies on specimens cut from bulk using electrical discharge machining (EDM) or abrasive water jet. But both HELIX and SPHINX need polished surfaces, requiring extra sample preparation and probably some ability to inspect the surfaces prior to testing. In addition, we note that processes such as rolling and forging at scale typically involve levels of noise, vibration, heat, and general dirtiness that are incompatible with laser optics and other precision characterization instrumentation. Thus, it is likely that automated laboratories for characterization (such as \aimdl) and fabrication/processing will need to be physically separated, though possibly linked by robotic sample transfer systems.

\section{Summary}
\aimdl is a significant step forward in high-throughput characterization of bulk structural materials. The two bespoke instruments---HELIX for laser-driven shock studies and MAXIMA for microstructural characterization---each achieve sample throughput that improves on conventional techniques by several orders of magnitude. SPHINX shows that even complex commercial instruments can be integrated into a fully automated robotic laboratory. Successfully integrating multiple instruments requires a reliable and robust automation system capable of flexibly handling large numbers of samples. Finally, data obtained at this scale can only be truly useful with a data architecture for streaming data autonomously, executing complex workflows, and making data available in forms that are readily interpretable by both human and agentic AI users.

While computational and data-driven modeling are not the primary focus of this paper, they represent a growing component of \aimdl. As the lab continues to mature, these modeling capabilities will play an increasingly important role in understanding and predicting the links between material chemistry and microstructure to performance under high-temperature and high-rate conditions. Integrating predictive modeling with the lab will ultimately accelerate closed-loop materials discovery by helping to guide experiments toward the most promising regions of the materials design space.

%% file: 12-endmatter.tex
\section{Acknowledgements}
\label{sec:acknowledgements}
We gratefully acknowledge everyone who has contributed to the design, construction, and development of AIMD-L and its associated instruments, data infrastructure, and other capabilities. 

Research was accomplished through projects sponsored by the Army Research Laboratory under Cooperative Agreement Numbers W911NF-22-2-0014 and W911NF-23-2-0062. The views and conclusions contained in this document are those of the authors and should not be interpreted as representing the official policies, either expressed or implied, of the Army Research Laboratory or the U.S. Government. The U.S. Government is authorized to reproduce and distribute reprints for Government purposes notwithstanding any copyright notation herein. Additional support from the Johns Hopkins Whiting School of Engineering is gratefully acknowledged.

\section{Declaration of interests}
The authors declare no competing interests.